\documentclass[iop,tighten]{emulateapj}

\usepackage{subfigure}
\def\simgt{\lower 2pt \hbox{$\, \buildrel {\scriptstyle >}\over{\scriptstyle \sim}\,$}}
\def\simlt{\lower 2pt \hbox{$\, \buildrel {\scriptstyle <}\over{\scriptstyle \sim}\,$}}

\begin{document}

\title{The role of environment in low-level active galactic nucleus
activity: \\ no evidence for cluster enhancement}

\author{Brendan~Miller,$^{1}$ Elena~Gallo,$^{1}$ Tommaso~Treu,$^{2}$
and Jong-Hak~Woo$^{3}$}

\footnotetext[1]{Department of Astronomy, University of Michigan, Ann
Arbor, MI 48109, USA}

\footnotetext[2]{Physics Department, University of California, Santa
Barbara, CA 93106, USA}

\footnotetext[3]{Astronomy Program, Department of Physics and
Astronomy, Seoul National University, Seoul, Republic of Korea}

\begin{abstract}

We use the AMUSE-Virgo and AMUSE-Field surveys for nuclear
\hbox{X-ray} emission in early-type galaxies to conduct a controlled
comparison of low-level supermassive black hole activity within
cluster and field spheroids. While both the Virgo and the Field
samples feature highly sub-Eddington \hbox{X-ray} luminosities
($L_{\rm X}/L_{\rm Edd}$ between $\sim10^{-8}-10^{-4}$), we find that
after accounting for the influence of host galaxy stellar mass, the
field early-type galaxies tend toward marginally greater
($0.38\pm0.14$ dex) nuclear \hbox{X-ray} luminosities, at a given
black hole mass, than their cluster counterparts. This trend is
qualitatively consistent with the field black holes having access to a
greater reservoir of fuel, plausibly in the form of cold gas located
near the nucleus. We are able to rule out at high confidence the
alternative of enhanced \hbox{X-ray} activity within
clusters. Presuming nuclear \hbox{X-ray} emission correlates with the
total energy and momentum output of these weakly accreting black
holes, this indicates that low-level active galactic nucleus feedback
is not generally stronger within typical cluster galaxies than in the
field. These results confirm that for most cluster early-type galaxies
(i.e., excluding brightest cluster galaxies) direct environmental
effects, such as gas stripping, are more relevant in quenching star
formation.

\end{abstract}


\section{Introduction}

The gas content and stellar populations of early-type galaxies are
observed to depend upon their large-scale surroundings. Field
early-type galaxies typically contain more cold (\ion{H}{1}) gas,
sometimes alongside young stellar populations (Oosterloo et
al.~2010). Significant star formation in high density environments
occurs primarily between $3<z<5$, whereas in low density environments
it persists to $1<z<2$ or even lower for low-mass galaxies (Thomas et
al.~2005; Treu et al.~2005b; Gobat et al.~2008). Proposed mechanisms
of inhibiting star formation within clusters include gas removal
(e.g., starvation through ram pressure stripping, tidal stripping,
thermal evaporation, or other possibilities; Treu et al.~2003; Moran
et al.~2005; and references therein) or morphological quenching (i.e.,
stabilization of a gas disk through the build-up of a stellar
spheroid; Martig et al.~2009). Such processes would operate only at
low efficiency within the field.

Another potential mechanism for terminating star formation is feedback
from an active galactic nucleus (AGN). Radiation from an efficiently
accreting supermassive black hole can launch winds or directly heat
the surrounding gas (e.g., Ciotti \& Ostriker 2007; Proga et
al.~2010). The duration of such ``quasar mode'' highly-luminous AGN
activity is only $\simlt10^{8}$~yr (Yu \& Tremaine 2002), outside of
which black hole feeding is highly sub-Eddington. Numerical and
observational work indicates that mechanical feedback may be of
persistent importance even in weakly accreting systems, in the form of
outflows (Blandford \& Begelman 1999; Pellegrini et al.~2012) or jets
(Falcke et al.~2004). The inclusion in simulations of ``radio mode''
feedback for AGNs at the center of rich groups or clusters
successfully suppresses star formation and reproduces the red colors
of large ellipticals (Croton et al.~2006; Merloni \& Heinz 2007;
Khalatyan et al.~2008). Brightest cluster galaxies (BCGs) often
display direct evidence of mechanical feedback in the form of radio
jets and inflated bubbles displacing the hot \hbox{X-ray}-emitting
intracluster medium (ICM) (e.g., McNamara \& Nulsen 2007). At least in
luminous early-type galaxies, the rate of accretion onto the
supermassive black hole is well correlated with the emerging jet power
(e.g., Allen et al.~2006; Balmaverde et al.~2008).

The frequency of AGN activity, within both early and late-type
galaxies, may itself depend on local galaxy density, but
interpretation is complicated by the sometimes non-overlapping nature
of various observational activity indicators. The fractional rate of
emission-line galaxies is higher in the field, to a degree exceeding
differences in morphological distributions (Dressler et al.~1985). On
the other hand, the fractional rate for AGNs selected by \hbox{X-ray}
luminosities is similar between field and cluster samples (Martini et
al.~2007; Haggard et al.~2010). Position within the cluster may also
play a role (Gavazzi et al.~2011), although Atlee et al.~(2011) find
the intracluster radial distribution of \hbox{X-ray} or IR-selected
AGNs to be consistent with that of non-AGNs (while noting luminous
\hbox{X-ray} AGNs may be more centrally concentrated). Within relaxed
clusters, Ruderman \& Ebeling (2005) find an excess of \hbox{X-ray}
point sources both peaked within the central regions and more broadly
distributed near the virial radius, which they attribute to black hole
activity triggered by interaction with the BCG and by mergers,
respectively.

We aim to assess the incidence and magnitude of low-level supermassive
black hole activity within representative field versus cluster
early-type galaxies, and to determine whether star formation in
ordinary cluster early-type galaxies is likely primarily
AGN-quenched. We characterize activity using nuclear \hbox{X-ray}
emission, which directly measures high-energy accretion-linked
radiative output but more importantly serves as a plausible proxy for
mechanical feedback (Allen et al.~2006; Balmaverde et al.~2008). Our
samples are the AMUSE\footnote{AMUSE: AGN Multiwavelength Survey of
Early-Type Galaxies}-Virgo and AMUSE-Field surveys. Together, these
target 203 optically-selected local early-type galaxies, with both
surveys centered around Large {\it Chandra\/} Programs of ACIS-S3
snapshot (3--15~ks) observations (Virgo: ID 08900784, 454~ks, PI Treu;
Field: ID 11620915, 479~ks, PI Gallo), supplemented with deeper
archival {\it Chandra\/} coverage. The sample selection, data
reduction and analysis, and nuclear \hbox{X-ray} properties for the
AMUSE-Virgo and AMUSE-Field surveys are presented in Gallo et
al.~(2008, 2010; hereafter G08, G10) and Miller et al.~(2011;
hereafter M11), respectively, from which we obtain values of $L_{\rm
X}$, $M_{\rm BH}$, and $M_{\rm star}$ (the nuclear 0.3--10~keV
\hbox{X-ray} luminosity,\footnote{We assume no intrinsic absorption of
the nucleus in these early-type galaxies. $L_{\rm X}$ would be
$\sim$0.1 (0.5) dex greater for an intrinsic column of 10$^{21}$
(10$^{22}$)~cm$^{-2}$; the former value may be typical of late-type
galaxies (Bogd{\'a}n \& Gilfanov~2011).} black hole mass, and galaxy
stellar mass, which have units of erg~s$^{-1}$, $M_{\odot}$, and
$M_{\odot}$, respectively). We emphasize that the AMUSE samples are
unbiased with respect to nuclear \hbox{X-ray} properties, and in fact
almost all of the objects have $L_{\rm X}<10^{41}$~erg~s$^{-1}$ and
$L_{\rm X}/L_{\rm Edd}<10^{-5}$, reaching luminosities well below
commonly utilized formal AGN classification limits. Throughout this
work, errors are quoted as 1$\sigma$.

\begin{figure}
\includegraphics[scale=0.66]{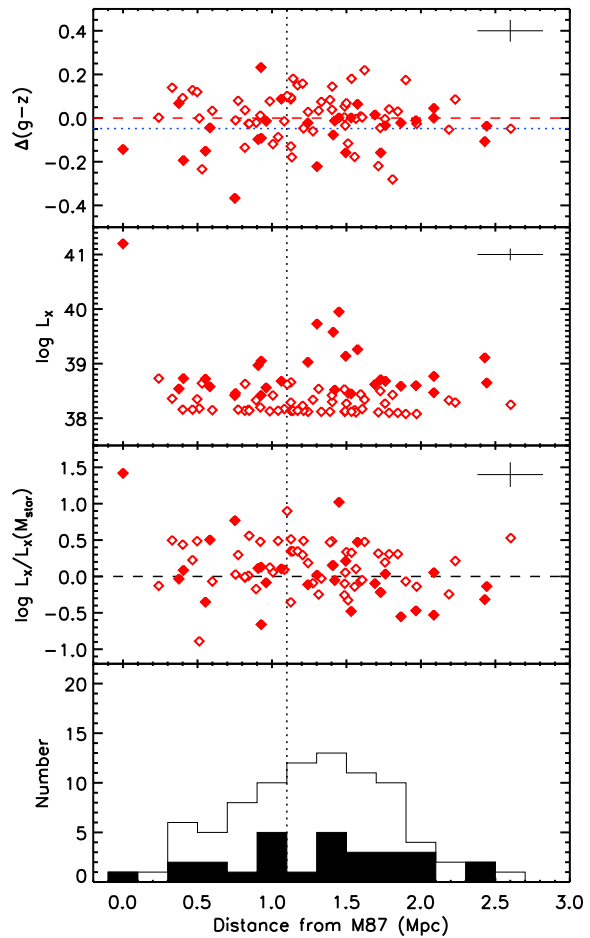} \figcaption{\small
Properties of the M87-associated Virgo galaxies versus distance from
M87. The vertical dotted line is $r_{\rm 200}$, and open symbols are
\hbox{X-ray} upper limits. {\it Top\/}: Relative color (dashed/dotted
lines are mean for Virgo/Field samples). {\it Middle \/}: \hbox{X-ray}
luminosity and residual luminosity. {\it Bottom \/}: \hbox{X-ray}
coverage (open) and detections (solid). Crosses show typical
uncertainties.}
\end{figure}

\section{Nuclear activity in field and cluster spheroids}

Before considering the overall characteristics of the Virgo sample, we
are motivated by the radial dependence of cluster potentials, gas
properties, and galactic densities to explore the relative colors and
\hbox{X-ray} properties of the included early-type galaxies as a
function of distance from M87.\footnote{Here only, Virgo galaxies
associated with the subcluster centered on M49 are excluded.}
Distances from M87 are calculated from the projected separation and
the Mei et al.~(2007) catalog of distance moduli. Figure~1 shows the
relative color,\footnote{We define relative color as
${\Delta}(g-z)=(g-z)-(0.21{\times}{\log{M_{\rm star}}}-0.83)$, based
on the Virgo red sequence. The Virgo and Field samples have mean
${\Delta}(g-z)=0.00$ and $-0.05$, respectively (see also Cassata et
al.~2007).} the \hbox{X-ray} luminosity, the residual \hbox{X-ray}
luminosity, and the detection fraction, as a function of distance from
M87. No strong trends are observed with any of these quantities within
the Virgo sample (cf.~$\S$1; Mart{\'{\i}}nez et al.~2010) and so it is
hereafter treated in its entirety.

\subsection{Comparison of AMUSE-Field and AMUSE-Virgo samples}

Table~1 contains the properties of the full AMUSE-Field and
AMUSE-Virgo samples as well as for the subset of objects for which
$M_{\rm BH}$ was calculated from the $M_{\rm BH}-\sigma$ relation
(G\"ultekin et al.~2009). For quantities derived from optical data
($\log{M_{\rm star}}$, $\log{M_{\rm BH}}$, and $\log{M_{\rm BH}/M_{\rm
star}}$), the 25th, 50th, and 75th percentiles are given as the value
for the nearest object in the sorted list. In addition to the nuclear
\hbox{X-ray} luminosity $L_{\rm X}$, we consider Eddington-scaled and
residual \hbox{X-ray} luminosities $\log{(L_{\rm X}/L_{\rm Edd})}$ and
$\log{L_{\rm X}/L_{\rm X}(M_{\rm star})}$, respectively, where $L_{\rm
X}(M_{\rm star})=38.36+0.71\times(\log{M_{\rm star}}-9.8)$ is the
best-fit relation determined for the Field sample from M11. For
quantities in Table~1 derived from \hbox{X-ray} data, the Kaplan Meier
distribution, incorporating upper limits, was determined using the
survival analysis package
ASURV\footnote{http://astrostatistics.psu.edu/statcodes/asurv}
(Lavalley et al.~1992); note, however, that at least the 25th
percentile values are dominated by censored points and so should be
taken as roughly indicative only. To ensure a uniform comparison, six
Field \hbox{X-ray} measurements (two detections) with $\log{L_{\rm
X}}<38.2$ are not used for this analysis; further, Virgo upper limits
with $\log{L_{\rm X}}<38.2$ are adjusted (by 0.15 dex) to match the
limiting Field sensitivity.

\begin{deluxetable*}{p{40pt}rrrrrcrrrrr}
\tablecaption{Sample Properties}
\tabletypesize{\footnotesize}
\tablewidth{17.1cm}

\tablehead{\colhead{Sample} & \colhead{$n$} & \colhead{Mean} &
\colhead{25th} & \colhead{50th} & \colhead{75th} & & \colhead{$n_{\rm
det}/n$} & \colhead{Mean} & \colhead{25th} & \colhead{50th} &
\colhead{75th}}

\startdata   
\multicolumn{6}{c}{$\log{(M_{\rm star}/M_{\odot})}$} &~~~& \multicolumn{5}{c}{$\log{(L_{\rm X}/10^{37})}$} \\
Field                  &  103 &   9.66$\pm$0.12  &  8.56  &  9.75  & 10.84 &  &  50/97 &  1.71$\pm$0.07 &  0.63 & 1.37 & 2.02  \\	       
~~With $\sigma$        &   61 &  10.54$\pm$0.09  & 10.02  & 10.70  & 11.10 &  &  46/57 &  2.01$\pm$0.09 &  1.52 & 1.92 & 2.33 \\[+3pt]   
Virgo                  &  100 &   9.90$\pm$0.08  &  9.20  &  9.80  & 10.40 &  & 32/100 &  1.43$\pm$0.05 &  0.46 & 0.91 & 1.55  \\       
~~With $\sigma$        &   54 &  10.48$\pm$0.09  & 10.10  & 10.40  & 10.80 &  &  28/54 &  1.60$\pm$0.08 &  0.73 & 1.44 & 1.72  \\[+4pt] 

\multicolumn{6}{c}{$\log{(M_{\rm BH}/M_{\odot})}$} &~~~& \multicolumn{5}{c}{$\log{(L_{\rm X}/L_{\rm Edd})}$} \\
Field                  &  103 &   6.90$\pm$0.12  &  5.75  &  6.79  &  7.96 &  &  50/97 &  $-$6.96$\pm$0.10 & $-$7.50 & $-$7.05 & $-$6.49  \\	     
~~With $\sigma$        &   61 &   7.66$\pm$0.12  &  7.25  &  7.89  &  8.40 &  &  46/57 &  $-$6.99$\pm$0.11 & $-$7.50 & $-$7.07 & $-$6.57  \\[+3pt]   
Virgo                  &  100 &   6.98$\pm$0.09  &  6.28  &  6.79  &  7.70 &  & 32/100 &  $-$7.42$\pm$0.12 & $-$7.78 & $-$7.38 & $-$6.87  \\ 	     
~~With $\sigma$        &   54 &   7.49$\pm$0.14  &  7.04  &  7.52  &  8.20 &  &  28/54 &  $-$7.45$\pm$0.13 & $-$7.79 & $-$7.43 & $-$7.02  \\[+4pt]       

\multicolumn{6}{c}{$\log{(M_{\rm BH}/M_{\rm star})}$} &~~~& \multicolumn{5}{c}{$\log{(L_{\rm X}/L_{\rm X}(M_{\rm star}))}$} \\  
Field                  &  103 &  $-$2.76$\pm$0.04  & $-$2.90  & $-$2.75  & $-$2.53 &  &  50/97 &   0.05$\pm$0.08 & $-$0.36 &  0.05 &  0.38  \\ 	        
~~With $\sigma$        &   61 &  $-$2.88$\pm$0.06  & $-$3.07  & $-$2.87  & $-$2.64 &  &  46/57 &   0.03$\pm$0.08 & $-$0.38 &  0.04 &  0.34  \\[+3pt]          
Virgo                  &  100 &  $-$2.92$\pm$0.04  & $-$3.01  & $-$2.86  & $-$2.72 &  & 32/100 &  $-$0.46$\pm$0.10 & $-$1.12 & $-$0.45 & $-$0.03  \\                 
~~With $\sigma$        &   54 &  $-$2.99$\pm$0.07  & $-$3.21  & $-$2.94  & $-$2.62 &  &  28/54 &  $-$0.45$\pm$0.11 & $-$1.13 & $-$0.45 & $-$0.04  \\[-15pt]         
\enddata

\tablecomments{Quantities are defined in $\S$1 or
$\S$2.1. \hbox{X-ray} distributions for the Field sample are
restricted to $\log{L_{\rm X}}>38.2$ and Virgo limits with
$\log{L_{\rm X}}<38.2$ have been adjusted by 0.15 dex to match the
Field sensitivity ($\S$2.1). Kaplan-Meier values for the \hbox{X-ray}
distributions are slightly biased because the first upper limit is
treated as a detection.}

\end{deluxetable*}

\begin{figure}
\includegraphics[scale=0.51]{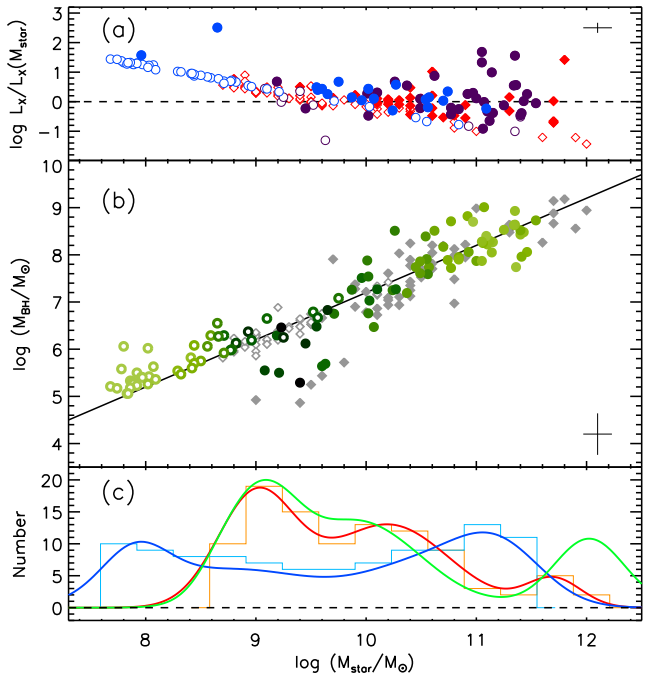} \figcaption{\small {\it
Top\/}: Residual \hbox{X-ray} luminosities (see $\S$2.1) for the Field
(blue circles: snapshot; purple circles: archival) and Virgo (red
diamonds) samples. Open symbols are upper limits. {\it Middle\/}:
$M_{\rm BH}$ versus $M_{\rm star}$ for the Field (green circles) and
Virgo (gray diamonds) samples. The line indicates the median
$\log{(M_{\rm BH}/M_{\rm star})}=-2.8$. Filled symbols have $M_{\rm
BH}$ calculated from $\sigma$. The tint of the Field points indicates
in what fraction of weighted subsamples they are included. {\it
Bottom\/}: Histogram of $M_{\rm star}$ distribution for the Field
(cyan) and Virgo (orange) surveys, with overplotted representation by
four and three Gaussians, respectively. The green line is the ratio of
the Virgo to Field summed Gaussians (arbitrary normalization), which
is the weighting function used to draw Field subsamples.}
\end{figure}

The \hbox{X-ray} detection fraction is higher for the Field sample
compared to Virgo (50$\pm$7\% versus 32$\pm$6\%, with 1$\sigma$
Poisson errors), but within the shorter snapshot exposures, the rates
are closer (31$\pm$7\% versus 24$\pm$5\%). However, the distribution
of \hbox{X-ray} luminosities also tends toward higher values within
the Field sample (Table~1). At face value, the percentage of objects
with $L_{\rm X}\ge10^{39}$~erg~s$^{-1}$ is significantly greater in
the Field sample (25$\pm$5\% versus 10$\pm$3\%). This holds also for
residual \hbox{X-ray} luminosity: expressed as a percentage of the
full samples, the number of detected galaxies with $\log{L_{\rm
X}/L_{\rm X}(M_{\rm star})}>0.4$ is 18$\pm$4\% for the Field versus
5$\pm$2\% for Virgo (similar results apply for $0.0<\log{L_{\rm
X}/L_{\rm X}(M_{\rm star})}<0.5$; Figure~2a). Incorporating upper
limits (with the caveats noted above), not only the 75th percentiles,
but also the 25th and 50th percentiles for \hbox{X-ray} luminosity
(also Eddington-scaled or residual), are modestly enhanced by
$\sim$0.2--0.5 dex in the Field relative to the Virgo samples.

We desire to compare the functional dependence of $L_{\rm X}(M_{\rm
star})$ in the Field versus Virgo samples. However, the distribution
of $\log{M_{\rm star}}$ for the full Field sample is inconsistent
[Kolmogorov-Smirnov (KS) test probability $p<0.001$] with that of the
Virgo sample. As both star formation and nuclear activity are strong
functions of stellar mass as well as environment (e.g., Treu et
al.~2005a; Yee et al.~2005), we control the Field-versus-Virgo
comparison for $M_{\rm star}$ as described next.

\begin{deluxetable}{p{68pt}ccrrr}
\tablecaption{Correlations with X-ray luminosity}
\tabletypesize{\footnotesize}
\tablewidth{8.7cm}

\tablehead{ & \multicolumn{5}{c}{$(\log{L_{\rm
 X}}-38)=A+B{\times}(\log{M_{\rm BH}}-8)$} \\
 \colhead{Sample\tablenotemark{a}} & \colhead{$n$} & \colhead{$n_{\rm
 det}$} & \colhead{$A$} & \colhead{$B$} & \colhead{${\sigma}_{\rm 0}$}}

\startdata   
Field (full) &   97 & 50 &$1.05^{+0.10}_{-0.10}$  & $0.61^{+0.16}_{-0.13}$  & $0.61^{+0.08}_{-0.07}$     \\
Virgo (full) &  100 & 32 &$0.67^{+0.09}_{-0.10}$  & $0.62^{+0.16}_{-0.14}$  & $0.49^{+0.07}_{-0.07}$     \\[+7pt]
Field (with $\sigma$)&   57 & 46 &$1.07^{+0.11}_{-0.12}$  & $0.44^{+0.15}_{-0.13}$  & $0.62^{+0.09}_{-0.08}$   \\
Virgo (with $\sigma$)&   54 & 28 &$0.67^{+0.11}_{-0.13}$  & $0.54^{+0.17}_{-0.15}$  & $0.54^{+0.11}_{-0.09}$   \\[+7pt]
Field (LMXB cor) &   97 & 45 &$1.04^{+0.10}_{-0.11}$  & $0.67^{+0.18}_{-0.14}$  & $0.60^{+0.08}_{-0.07}$  \\
Virgo (LMXB cor) &  100 & 28 &$0.63^{+0.10}_{-0.10}$  & $0.64^{+0.18}_{-0.14}$  & $0.50^{+0.08}_{-0.06}$  \\[+7pt]
Field (low lum) &   90 & 43 &$0.90^{+0.08}_{-0.08}$  & $0.42^{+0.11}_{-0.10}$  & $0.35^{+0.07}_{-0.05}$   \\
Virgo (low lum) &   99 & 31 &$0.64^{+0.09}_{-0.09}$  & $0.47^{+0.13}_{-0.11}$  & $0.38^{+0.08}_{-0.07}$   \\[+7pt]
Field (X-ray det)&   50 & 50 &$1.24^{+0.11}_{-0.11}$  & $0.30^{+0.13}_{-0.13}$  & $0.57^{+0.08}_{-0.06}$   \\
Virgo (X-ray det)&   32 & 32 &$0.99^{+0.12}_{-0.11}$  & $0.45^{+0.15}_{-0.15}$  & $0.42^{+0.08}_{-0.08}$   \\[+7pt]
Field (weighted)&   45 & 24$^{+2}_{-1}$  &$0.99^{+0.08}_{-0.07}$  & $0.51^{+0.05}_{-0.08}$  & $0.53^{+0.09}_{-0.03}$   \\[-15pt]  
\enddata

\tablecomments{Fitting of $L_{\rm X}(M_{\rm BH})$ is described in
$\S$2.3. The reported parameters are medians of the posterior
distributions, and the quoted errors correspond to 1$\sigma$ for one
parameter of interest. }

\tablenotetext{a}{The various samples are defined as follows: {\it
full\/}: all objects; {\it with $\sigma$\/}: only objects for which
$M_{\rm BH}$ was calculated from a high-quality measurement of
$\sigma$; {\it LMXB cor\/}: $L_{\rm X}$ changed to an upper limit for
those objects for which the probability of LMXB contamination is
non-negligible; {\it low lum\/}: only objects satisfying $L_{\rm
X}<10^{40}$ erg~s$^{-1}$; {\it X-ray det\/}: only objects with X-ray
detections. }

\end{deluxetable}

\subsection{Controlling for host stellar mass}

It is unlikely that the modest tendency toward greater \hbox{X-ray}
luminosities within the Field sample is due to the differing $M_{\rm
star}$ distributions: (i) the tail to low stellar masses that
primarily distinguishes the Field sample contains objects with
generally lower, not higher, \hbox{X-ray} luminosities; (ii) the
modest enhancement persists across the full range of considered
stellar masses (Figure~2a); (iii) the subsets of the Field and Virgo
samples for which $\sigma$ was used to calculate $M_{\rm BH}$ do have
consistent distributions of $M_{\rm star}$ (Figure~2b; these are
primarily the brighter galaxies), and here too the Field subset has
modestly greater \hbox{X-ray} luminosity (also Eddington-scaled or
residual).

We verify that $L_{\rm X}$, $M_{\rm BH}$, and $M_{\rm star}$ are all
mutually significantly correlated using the
method\footnote{http://www.astrostatistics.psu.edu/statcodes/cens\_tau}
of Akritas \& Siebert (1996; cf.~Kelly et al.~2007), which
incorporates censoring into the calculation of Kendall's partial
${\tau}_{12,3}$. For the Field sample, with (1) $\log{M_{\rm BH}}$ and
(2) $\log{L_{\rm X}}$ controlling for (3) $\log{M_{\rm star}}$, the
individual coefficients are ${\tau}_{\rm 12}=0.49$, ${\tau}_{\rm
13}=0.77$, and ${\tau}_{\rm 23}=0.50$ (all variables are correlated),
with ${\tau}_{\rm 12,3}=0.19\pm0.04$; the null hypothesis of zero
partial correlation is rejected. The $M_{\rm star}-M_{\rm BH}$
correlation would introduce degeneracy were both variables included,
while the $L_{\rm X}-M_{\rm star}$ correlation suggests $L_{\rm
X}(M_{\rm BH})$ is best compared across groups possessing consistent
$M_{\rm star}$ distributions.

In addition to analyzing the $L_{\rm X}(M_{\rm BH})$ relation for the
full Field and Virgo samples, we conduct two weighted comparisons. As
mentioned, the subsets of early-type galaxies for which $M_{\rm BH}$
is calculated from $\sigma$ already have consistent distributions of
$M_{\rm star}$. We also draw 21 random subsamples from the Field
survey weighted to correspond to the Virgo $M_{\rm star}$ distribution
(using the ratio of the Virgo to Field $M_{\rm star}$ histograms;
Figure~2c). This importance sampling is conducted without replacement,
placing an effective limit on the subsample size; we use $n=45$ as
only $\sim$5\% of such subsamples have $\log{M_{\rm star}}$
distributions inconsistent (KS $p<0.05$) with Virgo. Of the 97 Field
galaxies with $\log{L_{\rm X}}>38.2$, 16 (all with $\log{M_{\rm
star}}<8.1$) are not included in any of the 21 subsamples, while 48
[33] are included in (1 to 14)/21 [(15 to 21)/21] of the subsamples.

\subsection{Field versus cluster $L_{\rm X}(M_{\rm BH})$ correlations}

We parameterize the dependence of nuclear \hbox{X-ray} luminosity upon
black hole mass as $(\log{L_{\rm X}}-38)=A+B{\times}(\log{M_{\rm
BH}}-8)$, following the Bayesian methodology of G10 for
fitting. Uncertainties on $L_{\rm X}$ and $M_{\rm BH}$ are 0.11 and
0.44 dex, respectively, with Gaussian likelihood functions on the log
quantities (except uniform probability below upper limits on $L_{\rm
X}$). Rotational invariance is enforced on the power-law index $B$,
and the prior distribution of $M_{\rm BH}$ is taken to be
log-uniform. A Markov Chain Monte Carlo sampler is used to explore the
$\{A, B, {\sigma}_{\rm 0}\}$ parameter space. We use the median of
9000 random draws from the posterior distribution as the most likely
parameter value, with estimated 1$\sigma$ errors reported as the 16th
and 84th percentiles. We have improved the treatment of upper limits
over that in G10 and here use a more recent $M_{\rm BH}-\sigma$
relation to recalculate their black hole masses, and so we give
updated fits for the Virgo sample as well as new results for the Field
sample in Table~2, illustrated in Figure~3.

\begin{figure}
\includegraphics[scale=0.45]{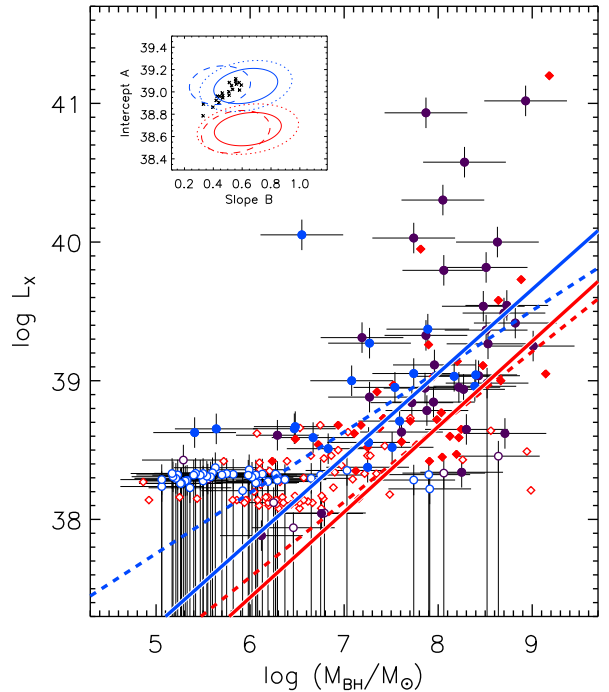} \figcaption{\small $L_{\rm X}$
versus $M_{\rm BH}$ for the Field and Virgo samples, with symbols
coded as in Figure~2a. The blue/red lines are the Field/Virgo fits to
the model $(\log{L_{\rm X}}-38)=A+B{\times}(\log{M_{\rm BH}}-8)$; the
solid lines are for the full samples, while the dashed lines are for
the objects with $M_{\rm BH}$ calculated from $\sigma$. The inset
shows joint 68\% (solid) and 90\% (dotted) confidence ellipses for the
full samples, and joint 68\% (dashed) confidence ellipses for the
$\sigma$ subsets. The best-fit parameters for the 21 $M_{\rm
star}$--weighted Field subsamples are also marked.}
\end{figure}

The best-fit relation for the full Field sample is $(\log{L_{\rm
X}}-38)=(1.05\pm0.10)+(0.61\pm0.15){\times}(\log{M_{\rm BH}}-8)$, with
intrinsic scatter of $0.61\pm0.08$ dex.\footnote{For reference, the
coefficients for this fit performed with the IDL Bayesian code of
Kelly~(2007) are $A=1.10\pm0.10$, $B=0.75\pm0.11$, and ${\sigma}_{\rm
0}=0.70\pm0.09$; the differing methodologies provide results
consistent within the errors.} Fitting simulated \hbox{X-ray}
luminosities for each Field $M_{\rm BH}$ point (randomly distributed
with ${\sigma}_{\rm 0}=0.61$ about the best-fit relation and with
values of $\log{L_{\rm X}}<38.4$ translated into limits closely
scattered near the $L_{\rm X}$ sensitivity threshold) returns output
coefficients in close agreement (${\Delta}A\simlt$10\%,
${\Delta}B\simlt$7\%) with those input. Better data (e.g., dynamical
black hole masses) or alternative priors or methods might yield
somewhat different coefficients, but we emphasize it is the comparison
between the (identically fit) Field and Virgo relations that is of
interest here.

The best-fit slopes for the full Field and Virgo surveys are
consistent, $B\simeq0.6$. Consequently, the average Eddington-scaled
\hbox{X-ray} luminosity scales with black hole mass as
${\propto}M_{\rm BH}^{-0.4}$ (a ``downsizing'' effect as noted by
G10). However, the best-fit Field intercept exceeds that for Virgo by
$0.38\pm0.14$, consistent with the modest \hbox{X-ray} enhancement
discussed in $\S$2.1. The complementary question of whether the Virgo
early-type galaxies are systematically more X-ray active is robustly
answered in the negative; the probability that the Field intercept is
$>2.5\sigma$ lower than that for Virgo is $<10^{-6}$.

Fitting to the measured-$\sigma$ subsets produces similar results,
with a difference in intercept of 0.40$\pm$0.17. (Somewhat flatter
slopes may derive from the incompleteness of $\sigma$ measurements in
fainter galaxies, which tend to be \hbox{X-ray} limits.) Fitting most
of the $M_{\rm star}$--weighted Field subsamples gives results
consistent with those for the full Field sample; 20th, 50th, and 80th
percentiles for each parameter from these fits are given in
Table~2. Converting the handful of \hbox{X-ray} detections with a
non-negligible possibility of LMXB contamination to upper limits does
not significantly alter the results. If objects with $\log{L_{\rm
X}}>40$ erg~s$^{-1}$ are arbitrarily excluded, thereby removing even
weak Seyfert-level activity, the slopes flatten somewhat and the
difference between intercepts is $0.26\pm0.12$. Restricting
consideration to \hbox{X-ray} detections also flattens the slopes;
here the difference between intercepts is $0.25\pm0.16$. Joint
confidence regions for the full and $\sigma$ subset fits are plotted
inset to Figure~3, along with the best-fit parameters for the $M_{\rm
star}$--weighted Field subsamples.

\section{Discussion}

We briefly consider the implications of enhanced $L_{\rm X}$ in field
early-types. As both the Field and Virgo galaxies have highly
sub-Eddington luminosities ($10^{-8}<L_{\rm X}/L_{\rm Edd}<10^{-4}$),
their accretion structures should be physically similar (see, e.g.,
Soria et al.~2006 for discussion of inefficient flow models). One
obvious parameter that could link nuclear activity to the large-scale
environment is the fuel supply. Hot gas is subject to offsetting
effects in clusters, with ram pressure stripping countered by
accretion from the ICM and potential confinement of winds (Brown \&
Bregman 2000), but radio and optical observations establish cold gas
and younger stellar populations as more prevalent in field early-type
galaxies ($\S$1). Major tidal interactions, which are less frequent
within clusters due to the large galaxy velocity dispersions, could
help bring gas to the nucleus, although this is not required as
stochastic events may suffice for low-level fueling (as conjectured
for Seyferts; Hopkins \& Hernquist~2006). While accretion in local
early-type galaxies does not appear to be limited by gas supply (Soria
et al.~2006; Pellegrini 2010), a $\sim$2$\times$ larger infall rate
(small in absolute terms; e.g., Allen et al.~2006) could plausibly
produce the modest observed enhancements in $L_{\rm X}$ assuming
uniform efficiency and outflow fraction.

The Field sample contains galaxies in groups spanning a range in
richness, as well as triples, pairs, or isolated galaxies. Groups are
intermediate between field and clusters in terms of strength of
galaxy-medium interactions, but facilitate strong galaxy-galaxy
interactions because the bulk speeds of the galaxies are lower
(somewhat similar to cluster outskirts). An apparent smooth decrease
in scaled \hbox{X-ray} luminosities from isolated to group to cluster
environments, albeit with large scatter, was interpreted by M11 as
tentative evidence of environmental modulation of supermassive black
hole fueling. The direct comparison between field and cluster galaxies
conducted in this work supports that possibility.

The robust result that these Virgo early-type galaxies are not
systematically more \hbox{X-ray} luminous than their Field
counterparts indicates that low-level black hole activity is not
generally stronger within typical cluster galaxies. This result is
supported by the insensitivity to environment of the fractional rate
of \hbox{X-ray}-identified AGNs (Martini et al.~2007; Haggard et
al.~2010). Presuming nuclear \hbox{X-ray} emission correlates with the
total energy and momentum output of these weakly accreting black holes
($\S$1), this implies low-level AGN feedback is not generally stronger
within typical cluster galaxies than in the field. This supports that
the older stellar populations in cluster early-type galaxies, and the
associated paucity of ongoing star formation, are not directly due to
recent black hole activity.

Possibly past AGN outbursts could have already suppressed star
formation and expelled gas from cluster galaxies, leaving no obvious
current link. Both star formation (Moran et al.~2005) and AGN activity
(Ruderman \& Ebeling 2005) may be triggered at infall near the cluster
virial radius, either through environmental or galaxy
interactions. The AGN fraction increases with redshift (e.g., Martini
et al.~2009; Haggard et al.~2010; Aird et al.~2011), and at $z\sim3$
may be significantly higher in clusters (Lehmer et al.~2009). However,
given that supermassive black hole activity is not generally greater
within clusters in the local universe, this scenario requires AGN
activity to be terminated or reduced more rapidly within
clusters. While various environmental effects may remove gas otherwise
available for AGN fueling in clusters, as described above, this
depletion also acts to quench star formation and so it is unclear that
past AGN activity is required for this task. Our findings are
consistent with other arguments that AGN activity is not the proximate
cause of star formation quenching, including the observed gradual
decline in remnant star formation with decreasing cluster radius which
led Moran et al.~(2005) to associate the cessation of star formation
with a slow-acting effect such as starvation (see also von der Linden
et al.~2010), and the similarity in the distribution of \hbox{X-ray}
luminosities for $9.5<\log{(M_{\rm star}/M_{\odot})}<12$ which led
Aird et al.~(2011)\footnote{Pellegrini (2010) also mention the large
scatter in \hbox{X-ray} luminosities above $M_{\rm
star}\simeq6\times10^{9} M_{\odot}$.} to state no evidence for
AGN-quenching.

We note BCGs occupy a privileged position centered within the cluster
potential and show direct evidence ($\S$1) of mechanical AGN
feedback. An additional complication is that BCGs in cool core
clusters do show recent star formation related to the cooling gas
(Hicks et al.~2010), again distinguishing them from typical cluster
galaxies. Our cluster sample is exclusive to Virgo, so we do not
investigate BCGs in general, but our conclusions about the importance
of environmental effects versus AGN feedback likely do not apply to
BCGs.

\acknowledgments

We thank Phil Marshall for code development and assistance with
fitting the $L_{\rm X}(M_{\rm BH})$ relations, and an anonymous
referee for comments that improved this work. T.~T.~acknowledges
support from the NSF through CAREER award NSF-0642621, and from the
Packard Foundation through a Packard Research
Fellowship. J.~H.~W.~acknowledges support by the Basic Science
Research Program through the National Research Foundation of Korea
funded by the Ministry of Education, Science and Technology
(2010-0021558). Support for this work was provided by NASA through
{\it Chandra\/} Awards Number 11620915 (AMUSE-Field) and 08900784
(AMUSE-Virgo) issued by the {\it Chandra\/} \hbox{X-ray} Observatory
Center.

\end{document}